\documentclass[twocolumn,aps]{revtex4}
\usepackage{amssymb}
\usepackage{epsfig}
\usepackage{psfig}
\usepackage{amsmath}

\newcommand {\be}{\begin{equation}}
\newcommand {\ee}{\end{equation}}
\newcommand {\bea}{\begin{eqnarray}}
\newcommand {\eea}{\end{eqnarray}}

\begin{document}

\title{Cosmological Constraints on a Dynamical Electron Mass}
\author{John D Barrow$^{1}$ and Jo\~{a}o Magueijo$^{2}$}
\affiliation{$^{1}$DAMTP, Centre for Mathematical Sciences, Cambridge University,
Wilberforce Rd., Cambridge CB3 0WA\\
$^{2}$Theoretical Physics, The Blackett Laboratory, Imperial College, Prince
Consort Rd., London SW7 2BZ}

\begin{abstract}
Motivated by recent astrophysical observations of quasar absorption systems,
we formulate a simple theory where the electron to proton mass ratio $\mu
=m_{e}/m_{p}$ is allowed to vary in space-time. In such a minimal theory
only the electron mass varies, with $\alpha $ and $m_{p}$ kept constant. We
find that changes in $\mu $ will be driven by the electronic energy density
after the electron mass threshold is crossed. Particle production in this
scenario is negligible. The cosmological constraints imposed by recent
astronomical observations are very weak, due to the low mass density in
electrons. Unlike in similar theories for spacetime variation of the fine
structure constant, the observational constraints on variations in $\mu $
imposed by the weak equivalence principle are much more stringent
constraints than those from quasar spectra. Any time-variation in the
electron-proton mass ratio must be less than one part in $10^{9}$since
redshifts $z\approx 1.$This is more than one thousand times smaller than
current spectroscopic sensitivities can achieve. Astronomically observable
variations in the electron-proton must therefore arise directly from effects
induced by varying fine structure 'constant' or by processes associated with
internal proton structure. We also place a new upper bound of $2\times
10^{-8}$ on any large-scale spatial variation of $\mu $ that is compatible
with the isotropy of the microwave background radiation.
\end{abstract}

\maketitle

%\title{Constraints on a Dynamical Electron Mass}

%\twocolumn[\hsize\textwidth\columnwidth\hsize\csname @twocolumnfalse\endcsname

%] \renewcommand{\thefootnote}{\arabic{footnote}} \setcounter{footnote}{0}

\section{Introduction}

This work is motivated by recent observational constraints~\cite{ubachs,
petit, tz} on variations of the electron-proton mass ratio $\mu =m_{e}/m_{p}$
which take advantage of new high-precision spectra to complement previous
investigations of $\mu $ in combination with the fine structure constant, $%
\alpha $, and the proton g-factor, $g_{p},$ \cite{cowie, thomp, foltz}. It
is important to relate these constraints to independent investigations on
varying $\alpha $ \cite{murphy, chand, quast}. The data reported in ref \cite%
{tz} from 21 cm and quasar UV spectra are drawn from absorption spectra in
the redshift range $0.24<z<2.04$ and lead to an observational constraint on
shifts in $x=g_{p}\alpha ^{2}\mu $ of $\Delta x/x=0.35\pm 1.09\times 10^{-5}$%
. As discussed in detail in ref \cite{tz}, using data on variations
permitted in $\alpha $ by Murphy et al \cite{murphy} and Chand et al \cite%
{chand} this results in constraints on variations in $\mu $ of $\Delta \mu
/\mu =1.5\pm 1.1\times 10^{-5}$ and $\Delta \mu /\mu =0.5\pm 1.1\times
10^{-5}$, respectively, since we expect that the variations in the $g$%
-factor $\Delta g_{p}/g_{p}\approx -0.1\Delta (m_{q}/\Lambda
_{QCD})/m_{q}/\Lambda _{QCD}$ will be negligible \cite{flam}. The question
we ask in this paper is: how should theorists react should a varying $\mu $
be observed but not a varying alpha? Could a simple framework accommodate
such a situation?

Theoretically there is a wide range of possible connections between $\alpha $
and $\mu $. In previously considered models variations in $\alpha $ lead to
variations in the electron mass (via the electron self-energy) and in the
proton mass (via the electrostatic energy contained inside a proton). These
are model dependent and in general quite complex to work out \cite{stein,
uzan,dine}. However it certainly seems the case that a varying alpha entails
a varying $\mu$.

However, superficially it should also possible to have a variation in $\mu $
without a variation in $\alpha $. The idea is simply to induce such
variations directly in the mass parameter for the electron. At least at tree
level, such variations would have no effect upon the mass of the proton, or
the value of $\alpha $. This constitutes the simplest varying $\mu $ theory
and in this paper we investigate some of its cosmological consequences
together with the constraints that can be placed by considering the allowed
level of violations of the weak equivalence theory. The analogous theory for
variation of $\alpha $ developed in ref. \cite{bsbm} produced observable
effects on quasar spectra at high redshift ($z\approx 1$) with accompanying
violations of the weak equivalence principle of order $10^{-13}$, an order
of magnitude below observational bounds. In contrast, if a varying $\mu $
exists at the astronomically observable level it result in violations of the
weak equivalence principle that are unacceptably large by a factor of more
than $10^{3}.$

\section{The formalism}

We shall use metric convention $+,-,-,-$ and set $\hslash =c=1$ throughout.
Consider the standard Dirac lagrangian 
\begin{equation}
\mathcal{L}_{\Psi }=\imath {\overline{\Psi }}\gamma ^{\mu }\partial _{\mu
}\Psi -m{\overline{\Psi }}\Psi
\end{equation}%
and let the electron mass be controlled by a \textquotedblleft
dilaton\textquotedblright\ field $\phi $ defined by $m=m_{0}\exp \phi$,
where $m_{0}$ is the current electron mass (so that $\phi =0$ today). The
minimal dynamics for $\phi $ may be set by the kinetic lagrangian 
\begin{equation}
\mathcal{L}_{\phi }={\frac{\omega }{2}}\partial _{\mu }\phi \partial ^{\mu
}\phi
\end{equation}%
where $\omega $ is a coupling constant. This is the minimal theory; we can
add a mass, a potential, or any other complication if required.

From this lagrangian we obtain a Dirac equation with variable electron (and
positron) mass 
\begin{equation}
(\imath \gamma ^{\mu }\partial _{\mu }-m)\Psi =0.
\end{equation}%
The dynamical equation for the logarithm of the mass ($\phi =\ln (m/m_{0})$)
is the driven wave equation 
\begin{equation}
\partial ^{2}\phi =-{\frac{m}{\omega }}{\overline{\Psi }}\Psi  \label{dynphi}
\end{equation}%
Performing a simple calculation we can derive the following result
concerning the driving term: the macroscopically averaged value of ${%
\overline{\Psi }}\Psi $ is negligible in the relativistic regime and is
given approximately by the electron and positron number density $%
(n_{e}+n_{p})$ in the non-relativistic regime. Thus, unlike the situation in
the simplest theories for variations in $\alpha $, \cite{bek,bsbm},
variations in $\mu $ do occur in the radiation era and start as soon as the
universe cools down below the electron rest-mass threshold.

The calculation is straightforward~\cite{itz}. We want to compute the
macroscopically averaged value of ${\overline{\Psi }}\Psi $: 
\begin{equation}
{\langle {\overline{\Psi }}\Psi \rangle }={\frac{1}{V\Delta t}}\int d^{3}xdt{%
\overline{\Psi }}\Psi \,.  \label{aver}
\end{equation}%
From the standard expansion 
\begin{equation}
\Psi (x)=\int \tilde{d^{3}p}\sum_{s=1}^{2}{\left( a_{s}(\mathbf{p})u_{s}(%
\mathbf{p})e^{-ip\cdot x}+b_{s}^{\dag }(\mathbf{p})v_{s}(\mathbf{p}%
)e^{ip\cdot x}\right) }
\end{equation}%
we arrive at 
\begin{eqnarray}
\int d^{3}x{\overline{\Psi }}\Psi &=&\int \tilde{d^{3}p}{\frac{m}{E}}%
\sum_{r}(a_{r}^{\dagger }(\mathbf{p})a_{r}(\mathbf{p})-b_{r}(\mathbf{p}%
)b_{r}^{\dagger }(\mathbf{p}))  \notag \\
&+&\sum_{rs}(a_{r}^{\dagger }(\mathbf{p})b_{s}^{\dagger }(-\mathbf{p}){%
\overline{u}}_{r}(\mathbf{p})v_{s}(-\mathbf{p}))e^{2iEt}  \notag \\
&+&\sum_{rs}(a_{r}(\mathbf{p})b_{s}(-\mathbf{p}){\overline{v}}_{r}(\mathbf{p}%
)u_{s}(-\mathbf{p}))e^{-2iEt}.
\end{eqnarray}%
The last two terms average to zero when integrated in time. The remaining
terms, upon normal ordering, lead to 
\begin{equation}
{\langle {\overline{\Psi }}\Psi \rangle }={\frac{1}{V}}\int d^{3}p{\frac{m}{%
E(p)}}(N_{e}(\mathbf{p})+N_{p}(\mathbf{p}))
\end{equation}%
Thus, in the relativistic regime $E\gg m$ and the driving term is
negligible. In the non-relativistic regime, on the other hand, (\ref{aver})
reduces to $n_{e}+n_{p}$.

This theory conserves lepton number. The term $e^{\phi }{\overline{\Psi }}%
\Psi $ leads to Feynman diagrams involving $\phi $ and the $\Psi $, but if
electrons are created then an equal number of positrons must also be
created. In general, a varying mass leads to particle production (see~\cite%
{morik}; and compare to the case of pair production in an electric field~%
\cite{itz}). In the non-relativistic regime, however, this is negligible as
shown below.

\section{Cosmological solutions}

We assume that the variations in $\phi $ that drive variations in $\mu $ are
small in the sense that they do not produce significant contributions to the
Friedmann-like equation governing the dynamics of the expansion scale factor
of the universe, $a(t)$. This assumption will be confirmed by the results
obtained. In a cosmological setting the equation for $\phi $ is 
\begin{equation}
\ddot{\phi}+3{\frac{\dot{a}}{a}}\dot{\phi}\approx -{\frac{m}{\omega }}%
(n_{e}+n_{p})\Theta (m-T)  \label{cosmophi}
\end{equation}%
where $\Theta $ is the theta function and $a(t)$ will be the scale factor
for a spatially flat Friedmann universe containing radiation, matter and
quintessence. The limiting case of no mass variations corresponds to the $%
\omega \rightarrow \infty $ limit. We first try to constrain this parameter
cosmologically. We write: $n_{e}\approx fn_{b}=f\eta s$, where $n_{L}$, $%
n_{b}$ and $s$ are the lepton number, baryon number and entropy densities, $%
f\approx 0.95$ under standard assumptions and $\eta =n_{b}/s$ must be of
order $10^{-10}$ for Big Bang nucleosynthesis to produce the observed light
element abundances.

Lepton number (and charge) conservation in this theory implies that $%
n_{L}\propto 1/a^{3}$. This imposes the requirement that 
\begin{equation}
\dot{n}_{e}+3{\frac{\dot{a}}{a}}n_{e}=P(t)=\dot{n}_{p}+3{\frac{\dot{a}}{a}}%
n_{p}  \label{enP}
\end{equation}%
where $P(t)$ is the particle production rate density. Energy conservation in
turn requires that 
\begin{equation}
\dot{\rho}_{\phi }+3{\frac{\dot{a}}{a}}(\rho _{\phi }+p_{\phi }) =-mn_{L}%
\dot{\phi} = -\dot{\rho}_{L}-3{\frac{\dot{a}}{a}}\rho _{L}
\end{equation}
where $p_{\phi }=\rho _{\phi }=\omega \dot{\phi}^{2}/2$, and (in the
non-relativistic regime) $\rho _{L}=m(n_{e}+n_{p})$, $p_{L}\approx 0$.
Combining energy and lepton-number conservation we therefore conclude that $%
P(t)=0$ and we have $n_{e}=B/a^{3}$ and $n_{p}=A/a^{3}$; $A,B$ constants.
Although the electron number is conserved the varying mass induces a
corresponding variation in the electron's energy density. This is supplied
(or absorbed) by the $\phi $ field so that the overall energy is conserved.

We now consider a number of approximate solutions with cosmological
implications. In general, we can write (\ref{cosmophi}) as: 
\begin{equation}
(\dot{\phi}a^{3}\dot{)}=-M\exp [\phi ]  \label{eqM}
\end{equation}%
with 
\begin{equation}
M={\frac{a^{3}n_{L}m_{0}}{\omega }}\approx {\frac{\rho _{e0}a_{0}^{3}}{%
\omega }.}  \label{emdef}
\end{equation}%
If the mass variations are small, as observations demand, we may set $%
e^{\phi }\approx 1$ in the right hand side of (\ref{eqM}) and integrate to
obtain 
\begin{equation}
\dot{\phi}a^{3}=-Mt+\tilde{A}  \label{approx}
\end{equation}%
where $\tilde{A}$ is an integration constant. In the dust era, we take $a%
\propto t^{2/3}$, and assume that the driven solution dominates \cite{note2};
hence $\tilde{A}\approx 0$, and so $\phi =-Mt_{0}^{2}\ln (t/t_{0})$, that is 
\begin{equation}
\phi ={\frac{3}{2}}Mt_{0}^{2}\ln (1+z)  \label{dustsol}
\end{equation}%
where $t_{0}$ is the present time and $z$ is the redshift corresponding to
comoving proper time $t$. But 
\begin{equation}
Mt_{0}^{2}={\frac{\Omega _{e}}{\omega }}\rho _{c}t_{0}^{2}={\frac{\Omega _{b}%
}{\omega }}{\frac{m_{e}}{m_{p}}}{\left( 1-{\frac{f_{He}}{2}}\right) }{\frac{1%
}{6\pi G},}
\end{equation}%
where $f_{He}$ is the helium-4 number fraction, and so the shift in $\mu $
between redshift $z$ and the present is given by 
\begin{equation}
{\frac{\Delta \mu }{\mu }}\approx \phi ={\frac{1}{4\pi G\omega }}\Omega
_{b}\mu {\left( 1-{\frac{f_{He}}{2}}\right) }\ln (1+z).  \label{delmumat}
\end{equation}%
The observational constraints from redshifts of order 1 \cite{ubachs, petit,
tz} therefore convert into a bound on $\omega $ the one free parameter
defining the theory and 
\begin{equation}
{\frac{|\Delta \mu |}{\mu }}<10^{-5}\rightarrow G|\omega |>0.2.
\label{bound1}
\end{equation}%
This is a very poor constraint, a fact to be expected because the electron
density is so low. It means that $\omega $ can be extremely 'small' --
slightly smaller even than the minimal scale of $E_{planck}^{2}=G^{-1}$.
Recall that the no-$m_{e}$-variation model has $\omega =\infty $ and current
astronomical observations at high redshift barely constrain this maximal
variation case. Similar bounds on the combination $\alpha ^{2}\mu $ arise by
introducing an additional scalar field to carry variations in $\alpha $ as
in \cite{bsbm} and the bounds on the time variation of this combination are
dominated by the constraints on changes in $\mu $ if the two scalar fields
are coupled only by gravity.

In the above derivation we have neglected the acceleration of the universe.
This will weaken the bound further, as acceleration stabilizes $\mu $. If we
solve (\ref{eqM}) in a flat background universe containing dust and a
cosmological constant, $\Lambda $, and scale factor $a^{3}\propto \sinh
^{2}[t\sqrt{3\Lambda }/2]$ then the driven solution becomes 
\begin{equation}
\phi =\phi _{\ast }\ +\frac{M\Omega _{b0}\ }{\Omega _{\Lambda 0}\ \lambda \ }%
t\coth \lambda t-\frac{M\Omega _{b0}\ }{\ \Omega _{\Lambda 0}\lambda ^{2}\ }%
\ln \sinh \lambda t  \label{philambda}
\end{equation}%
where $\lambda =\sqrt{3\Lambda }/2$ and as $t\rightarrow \infty $ we have $%
\phi \rightarrow $ constant which we neglect as small. Similar solutions
have been found in~\cite{vet,olive}. To a reasonable approximation $\phi $
evolves as (\ref{dustsol}) until the epoch $z_{\Lambda }=\sqrt[3]{7/3\text{
\ }}-1$ $\approx 0.3$ when the cosmological constant begins to dominate the
expansion and is approximately constant thereafter. So the bound (\ref%
{bound1}) will be weakened by a factor of order $0.3$. The variation of $%
\phi $ is turned off in a similar way whenever the dynamics are dominated by
a fluid with equation of state satisfying $\rho +3p<0$ (such as
quintessence) and this$\ $includes the case of $\ $negative curvature ($a=t$%
) also. We see that our distance from the future asymptotic value, $\phi
(t_{0})-\phi _{\ast }$, is also small.

In the radiation epoch, where $a=t^{1/2},$ we find a different solution of (%
\ref{eqM}): %\begin{equation}
$\dot{\phi}a^{3}=-M(t-t_{m})$, %\end{equation}%
where $t_{m}$ $\approx 100s$ is the time when the electron mass threshold is
crossed at $T\sim m_{e}$. This integrates to give 
\begin{equation}
\phi =C-2Mt_{0}^{2}{\frac{a}{a_{0}}}{\left( 1+{\frac{t_{m}}{t}}\right) ,}
\end{equation}%
where the constant $C$ should be chosen so that at equality $\phi =\phi
_{eq} $ matches the result obtained from the dust solution (\ref{delmumat}).
As $z_{m}\gg z_{eq}$ we have 
\begin{equation}
\phi =\phi _{eq}+2Mt_{0}^{2}{\left( {\frac{1}{z_{eq}}}-{\frac{1}{z}}\right) }
\label{radsol}
\end{equation}%
(there is a small correction if $z\approx z_{m}\approx 10^{9}$). We see that
most of the variation in $\mu $ in the radiation epoch happens near the
epoch of matter-radiation equality, so that the redshift of the electron
mass threshold, $z_{m}$, is numerically irrelevant.

We can now consider the total variation in $\mu $ as we go up in redshift
deep into the radiation epoch, $z\rightarrow \infty $. This is given by 
\begin{equation}
{\frac{\Delta \mu }{\mu }}=\phi _{\infty }=Mt_{0}^{2}{\left( {\frac{3}{2}}%
\ln z_{eq}+{\frac{2}{z_{eq}}}\right) .}  \label{total}
\end{equation}%
The first term contains the variation that happened in the matter epoch; the
second adds on the variation that happened in the radiation epoch. The first
term dominates, and so most of the variation in the electron mass in this
theory occur in the matter epoch. Thus we do not expect detailed
consideration of primordial nucleosynthesis at $z\sim 10^{9}-10^{10}$ to
constrain it further. Note that our approximation was based on neglecting
the contribution of the scalar-field energy density to the Friedmann
equation. This requires $\omega \dot{\phi}^{2}/t^{-2}$ to be bounded at
early times and this holds for our solution (\ref{total}).

This leaves us with a worrying possibility. In the above discussion we have
assumed that before $t_{m}\sim m_{e}^{-1}$ there are no variations in $m_{e}$
since the driving term is absent. But this need not necessarily be the case.
We could add to the driven solution any solution to the free equation, $\dot{%
\phi}a^{3}=-A$, where $A$ is an arbitrary constant. In the
radiation-dominated case \cite{note3} this would add to $\phi $ solution 
%\begin{equation}
$\phi _{free}=-2A/t^{1/2}$ %\end{equation}%
As long as $A$ is chosen so that $\phi $ variations are suitably small at
nucleosynthesis, this component cannot be ruled out. Its effect, however,
would be dramatic. It would imply that the electron becomes massless (or
infinitely massive) at the Big Bang. However, if this solution mode is
included our approximation of neglecting the back-reaction of the $\phi $
motion on the Friedmannian dynamics of the universe begins to fail. The
kinetic term for this mode is of order $\omega \dot{\phi}^{2}\approx \omega
A^{2}a^{-6}$ and will dominate the radiation plasma ($\rho \propto a^{-4}$)
in the Friedmann equation as $t\rightarrow 0,$ leading to scalar-dominated
background expansion with $a=t^{1/3}$. The resulting exact general solution
of (\ref{eqM}) is

\begin{equation*}
m=\exp [\phi ]=-\frac{2C^{2}}{Mt}\left( \frac{t}{T}\right) ^{\pm C}\frac{1}{%
[1-(t/T)^{\pm C}]^{2}}
\end{equation*}%
with $C\ $constant. With $C=2n+1,n\in 
%TCIMACRO{\U{2124} }%
%BeginExpansion
\mathbb{Z}
%EndExpansion
^{+},T=-\beta <0,$ we have $m\propto t^{2n}\rightarrow 0$ as $t\rightarrow
0. $

\section{The weak equivalence principle}

Stronger bounds on this theory are possible if we consider the observational
constraints on any local violations of the weak equivalence principle (WEP).
All varying constant theories imply the existence of a 'fifth force' but
this need not violate the equivalence principle. Varying $e$ theories, like
BSBM, which match the level of variations in $\alpha $ consistent with
quasar data of refs \cite{murphy} predict WEP violations within one order of
magnitude of current constraints \cite{bsbm} and there is a real prospect of
testing this prediction with the first generation of space-based tests of
the WEP. In the varying-$\mu $ theory introduced here the situation turns
out to be quite different.

Defining $\zeta_t =\rho _{e}/\rho $, test particles may be represented by 
\begin{equation}
\mathcal{L}(y)=-\int d\tau \;m((1-\zeta _{t})+\zeta _{t}e^{\phi })[-g_{\mu
\nu }\dot{x}^{\mu }\dot{x}^{\nu }]^{\frac{1}{2}}{\frac{\delta (x-y)}{\sqrt{-g%
}}}
\end{equation}%
where over-dots are derivatives with respect to the proper time $\tau $.
This leads to equations of motion: 
\begin{equation}
\ddot{x}^{\mu }+\Gamma _{\alpha \beta }^{\mu }\dot{x}^{\alpha }\dot{x}%
^{\beta }-{\frac{\zeta _{t}e^{\phi }}{(1-\zeta _{t})+\zeta _{t}e^{\phi }}}%
\partial ^{\mu }\phi =0
\end{equation}%
which in the non-relativistic limit (with $\zeta _{t}\ll 1$) reduce to 
\begin{equation}
{\frac{d^{2}x^{i}}{dt^{2}}}=-\nabla _{i}\Phi +\zeta _{t}\nabla _{i}\phi \;,
\end{equation}%
where $\Phi $ is the gravitational potential. Linearizing (\ref{dynphi})
around a spherically symmetric body with mass $M_{s}$ and $\zeta =\zeta _{s}$
we have 
\begin{equation}
\phi =-{\frac{\zeta _{s}}{4\pi \omega }}{\frac{GM_{s}}{r}}
\end{equation}%
so that we predict an anomalous acceleration equal to 
\begin{equation}
a={\frac{GM_{s}}{r^{2}}}{\left( 1+{\frac{\zeta _{s}\zeta _{t}}{G\omega \pi }}%
\right) }  \label{acc}
\end{equation}%
Violations of the WEP occur because $\zeta _{t}$ is substance dependent. For
two test bodies with $\zeta _{1}$ and $\zeta _{2}$ the E\"{o}tv\"{o}s
parameter is: 
\begin{equation}
\eta \equiv {\frac{2|a_{1}-a_{2}|}{a_{1}+a_{2}}}={\frac{\zeta _{s}|\zeta
_{1}-\zeta _{2}|}{G\omega \pi }.}
\end{equation}%
Since $\zeta _{s}\approx |\zeta _{1}-\zeta _{2}|=\mathcal{O}(10^{-4})$ then
in order to produce $\eta <\mathcal{O}(10^{-12})$ we need $G\omega >10^{3}$.
Similar bounds are obtained by considering the motion of the Earth-Moon
system due to their compositional differences, \cite{nord, bart} Under this
constraint on $\omega $ the largest tolerated variations in $\mu $ at
redshifts of order $1$ from eq. (\ref{delmumat}) are of the order

\begin{equation}
|\Delta \mu |/\mu <10^{-9}.  \label{wep}
\end{equation}

It is easy to see why the varying-$\mu $ theory is so much more strongly
constrained by tests of the WEP than the BSBM varying-$\alpha $ theory~\cite%
{bek,bsbm}. The latter is a varying-$e$ theory, driven by $E^{2}-B^{2}$. It
is possible that the dark matter is dominated by $E^{2}$ or $B^{2}$ field:
this enhances the cosmological variations permitting variations in alpha at $%
z\approx 1$ of order $10^{-5}$ without conflict with the WEP, as explained
in \cite{bsbm}. A similar structure for the theory proposed in this paper
would require the dark matter to be dominated by electrons, that is $\zeta
_{D}\approx 1$. This is not a possibility. Still one should bear in mind
this possible loophole, as well as the one raised in~\cite{bekwep}.

\section{Cosmological inhomogeneities}

Another astronomical consideration is that of the non-uniform matter
distribution in the universe. In line with usual practice we have considered
the evolution of $\mu $ in a spatially homogeneous universe. Since the real
universe exhibits significant inhomogeneity in the matter distribution on
sub-Gpc scales we expect there will be inhomogeneity in $\mu $ also. From
eqns. (\ref{radsol})-(\ref{total}) we see that the fact that the dominant
contributions to variations in $\mu $ come from changes at low redshifts in
the dust-dominated era. The largest effects arising from the effects of
growing density inhomogeneities will arise at these times. The leading-order
effect in (\ref{total}) will be inhomogeneity in $M$ which arises directly
from inhomogeneity in $\rho _{e}$. Once protoclusters separate out from the
expansion of the universe and virialise the time evolution of $\mu $ will
cease inside them while $\mu $ continues to decrease logarithmically in the
background universe until the expansion starts to accelerate \cite{space}.
This will create small spatial variation in the value of $\mu $ between
clusters and the background universe and we can solve (\ref{approx}) to
determine the evolution of $\phi $ and $\mu $ inside the over-dense region.
If we model a growing spherical density inhomogeneity in the dust era by a
closed Friedmann universe \cite{space, pea} with greater than average
density we obtain %
%\begin{equation*}
$\phi =-Mt_{0}^{2}\left[ \ln (t/t_{0})+\Gamma (\vec{x})(t/t_{0})^{2/3}\right]
$ %\end{equation*}%
where $\Gamma (\vec{x})$ is a positive function of the space coordinates
that fixes the amplitude of the growing perturbation mode in the linear
approximation. Hence, substituting for redshift

\begin{equation*}
\phi =\frac{3}{2}Mt_{0}^{2}\ln (1+z)-\frac{\Gamma Mt_{0}^{2}}{1+z}.
\end{equation*}%
To leading order, the spatial inhomogeneity in $\mu \ $is

\begin{equation*}
\frac{\delta \mu }{\mu }=\frac{\mu -\bar{\mu}}{\bar{\mu}}\approx
-Mt_{0}^{2}\Gamma (\vec{x})\left( \frac{t}{t_{0}}\right) ^{2/3}=\frac{%
-Mt_{0}^{2}\Gamma (\vec{x})\ }{\ 1+z}
\end{equation*}%
where $\bar{\mu}$ is the spatially uniform value of $\mu $ in the background
(where $\Gamma =0$). Therefore we expect inhomogeneities in $\mu $ to grow
in time at the same rate as inhomogeneities in the matter density which
drive them. There is therefore observational motivation to search for
spatial variations in $\mu $ and in $\dot{\mu}$. Since the dominant
evolution of the electron mass occurs in the dust-dominated era there
appears to be little effect on primordial nucleosynthesis. There will be
effects on the Thomson scattering of the microwave photons (as $\sigma
_{T}\sim m_{e}^{-2}$) but they are also expected to be unobservably small.

It is possible to place a strong bound on the magnitude of any spatial
variations in $\mu $ by noting that inhomogeneity in $\mu $ is driven by
inhomogeneity in the density $\rho _{e}$ and the latter will lead to metric
potential perturbations on the scale of the inhomogeneity which will produce
temperature anisotropies in the microwave background on large angular scales 
\cite{jdbspat}. If we assume that the inhomogeneity in the electron density
is approximately proportional to that in the matter distribution, so

\begin{equation*}
\frac{\delta \rho _{e}}{\rho _{e}}=\beta \frac{\delta \rho _{m}}{\rho _{m}}.
\end{equation*}%
then the spatial inhomogeneity in the electron-proton mass ratio, $\mu ,$ is

\begin{equation*}
\frac{\delta \mu }{\mu }\sim \frac{0.3}{G\omega }\frac{\delta \rho _{e}}{%
\rho _{e}}\left( \frac{L}{t}\right) ^{2}\sim \frac{0.9\beta }{G\omega }\frac{%
\Delta T}{T}\sim \ \frac{2\times 10^{-5}\beta }{G\omega }<2\times 10^{-8}
\end{equation*}%
Here, we have used the WEP lower bound on $G\omega $, \ref{wep} to improve
on the cosmological bound given in \cite{jdbspat} on spatial inhomogeneity
in the electron-proton mass ratio on large cosmological scales that have not
undergone gravitational binding and non-linear evolution.

\section{Conclusions}

In summary, we have formulated the first theory which describes the
cosmological evolution of the electron-proton mass ratio, $\mu $. Although
'limits' exist on the time variation this quantity in the literature and the
consequences of altering its constant value are well known \cite{BT, JB, CR,
MT, CH} there has been no self-consistent theoretical description of the
time-evolution of $\mu $ that is consistent with the conservation of energy
and momentum and cosmology. In this paper we have presented the first (and
simplest) such theory. It can be related to dilaton couplings beyond
tree-level found in string theory~\cite{poly}, to the covariant varying-$c$
model~\cite{covvsl}, or to the standard model with varying parameters (as
long as the variations envisaged here are due to a dynamic Yukawa coupling
rather than a shift in the Higgs VEV~\cite{dag}). The theory has appealing
features which result in it being relatively insensitive to uncertainties in
defining parameters and we have found that it is remarkably weakly
constrained by the high-precision astronomical data provided by observations
of quasar spectra. However, unlike the case of theories of varying $\alpha $%
, \cite{bsbm}, the theory is significantly more strongly constrained by the
data supporting the weak equivalence principle. The deviations from
composition-independent freefall under gravity place constraints upon
varying $\mu $ which are about $1000$ times stronger than the current quasar
bounds and indicate that direct effects from varying $\mu $ on atomic energy
levels (in particular to the observable combination $\alpha ^{2}\mu $) at
redshifts $O(1)$ will be unobservably small (at most $O(10^{-9})$) if the
variations in $\mu $ are driven by scalar-field variation of the electron
mass as assumed here. Thus if observational evidence arises for cosmological
time variations in $\mu $ we expect complementary variations in $\alpha $ to
exist from which the $\mu $ variations arise from radiative corrections or
electrostatic effects internal to the proton.%\begin{acknowledgement}

\textbf{Acknowledgements} We thank Dagny Kimberly, Doug Shaw and John Webb
for helpful discussions, and the referees for suggesting significant
extensions. %\end{acknowledgement}


\begin{thebibliography}{99}
\bibitem{note1} The sense of the variation in $\mu $ is free in a way that
contrasts with varying $\alpha $ theories. In the latter the sign of the
variation is controlled by the relative magnitude of magnetic and electric
field contributions to the coupling.
\bibitem{note2} The general asymptotic form for (\ref{eqM}) is $\phi =\phi
_{0}+At^{-1}-\ln [\ln [C+t^{q}]],$ where $q=M\exp [\phi _{0}]$ and $C$ is
constant. Dropping the free solution mode ($A=0$) this reduces to our
solution (\ref{dustsol}) in the appropriate $t$ limit.
\bibitem{note3} The general radiation solution has the form $\phi =\phi
_{0}+Et^{-1/2}-2\ln (C+Dt^{n})$ with $n=1/4$ as $t\rightarrow \infty $ but $%
n=1/2$ at the early times we are investigating so $\phi (t)$ is well
approximated by eq. ( \ref{radsol}).
\end{thebibliography}
\end{document}